\documentclass[runningheads]{llncs}

\usepackage{graphicx}
\usepackage{multirow}
\usepackage{eucal}
\usepackage{multicol}
\usepackage{booktabs}
\usepackage{subfig}
\usepackage{amsmath}
\usepackage{amsmath,amssymb,amsfonts}
\usepackage{algorithm}
\usepackage{textcomp}
\usepackage[table]{xcolor}
\definecolor{lightgray}{gray}{0.9}
\usepackage{booktabs}
\usepackage{lineno}
\usepackage[table]{xcolor}%
\usepackage{xcolor,soul}
\sethlcolor{lightgray}
\usepackage{gensymb}
\usepackage{amssymb}
\usepackage{color}
\usepackage[colorlinks=true,
            linkcolor=red,
            citecolor=blue]{hyperref}
\usepackage{indentfirst}
\usepackage[noend]{algpseudocode}
\def\algbackskip{\hskip-\ALG@thistlm}
\usepackage{caption}
\usepackage{helvet}
\usepackage{courier}
\usepackage{bm}
\usepackage{epsfig}
\usepackage{enumitem}
\usepackage{rotating}
\graphicspath{{fig/}}
\usepackage[T1]{fontenc}
\usepackage{booktabs}
\usepackage{subfig}
\usepackage{url,longtable,multirow,morefloats,stfloats, floatflt,cancel,tfrupee}
\makeatletter
\AtBeginDocument{\@ifpackageloaded{textcomp}{}{\usepackage{textcomp}}}
\makeatother
\usepackage{colortbl,xcolor}
\usepackage{pifont}
\usepackage[nointegrals]{wasysym}
\usepackage[format=plain,
            labelfont=it,
            textfont=it]{caption}
\makeatletter
\def\algbackskip{\hskip-\ALG@thistlm}

\@ifundefined{etal}{}{}

\usepackage{ifxetex}
\ifxetex\else\if@twocolumn\@ifpackageloaded{stfloats}{}{\usepackage{dblfloatfix}}\fi\fi

\graphicspath{{fig/}}
\begin{document}
\title{Domain-Adaptive 3D Medical Image Synthesis: \\ An Efficient Unsupervised Approach}

\author {Qingqiao Hu* \inst{1} \and 
    Hongwei Bran Li* \inst{2,3} \and 
    Jianguo Zhang$\dagger$ \inst{1}}

\institute{Research Institute of Trustworthy Autonomous System and Department of Computer Science and Engineering, Southern University of Science and Technology, Shenzhen, China \and 
Department of Computer Science, Technical University of Munich, Germany \and 
Department of Quantitative Biomedicine, University of Zurich, Switzerland
\email{Email: winstonqhu@gmail.com, 
hongwei.li@tum.de}}

\maketitle              %
\begin{abstract}
Medical image synthesis has attracted increasing attention because it could generate missing image data, improving diagnosis and benefits many downstream tasks. 
However, so far the developed synthesis model is not adaptive to unseen data distribution that presents domain shift, limiting its applicability in clinical routine. 
This work focuses on exploring domain adaptation (DA) of \emph{3D} image-to-image synthesis models. 
First, we highlight the technical difference in DA between classification, segmentation and synthesis models. Second, we present a novel efficient adaptation approach based on 2D variational autoencoder which approximates 3D distributions. Third, we present empirical studies on the effect of the amount of adaptation data and the key hyper-parameters. Our results show that the proposed approach can significantly improve the synthesis accuracy on unseen domains in a 3D setting. The code is publicly available at \hyperref[]{\url{https://github.com/WinstonHuTiger/2D_VAE_UDA_for_3D_sythesis}}

\end{abstract}

\let\thefootnote\relax\footnotetext{\textsuperscript{$\ast$} Q. Hu and H. Li made equal contributions to this work. \\
\textsuperscript{$\dagger$} J. Zhang is the corresponding author. 
}
\section{Introduction}
Medical image synthesis is drawing increasing attention in medical imaging, because it could generate missing image data, improving diagnosis and benefits many downstream tasks such as image segmentation \cite{thomas2022improving,conte2021generative,finck2020deep}. For example, missing modality is a common issue in multi-modal neuroimaging, e.g., due to motion in the acquisition process \cite{Conte2021-ek}. However, existing synthesis frameworks are mostly developed and evaluated on single-domain data (e.g., images from the same scanner) with limited consideration of model robustness when testing on unseen image domains with different distributions, e.g., images collected from another imaging scanner or acquisition protocols. Hence, domain adaptation is crucial for the real-world deployment in clinical routine. In particular, unsupervised domain adaptation (UDA) is more practical as it does not require additional expensive supervision to fine-tune the pre-trained model.

It should be noted that UDA of classification \cite{tzeng2017adversarial,ahn2020unsupervised,lenga2020continual} and segmentation \cite{dou2018unsupervised,kamnitsas2017unsupervised,perone2019unsupervised,liu2021s,karani2021test} models are well explored in recent years. 
For image segmentation models, the problem formulation is as follows. Given two different \emph{input domains} (i.e., source and target) with data $X$ and its distribution $P(X)$, $\mathcal{D}_{s}$=$\{X_{s}, P(X_{s})\}$, $\mathcal{D}_{t}$=$\{X_{t}, P(X_{t})\}$ and a shared \emph{output} space $\mathcal{Y}=\{Y\}$, a predictive model
$f(\cdot)$ which approximates $P(Y|X)$ trained on the source domain $\mathcal{D}_{s}$ is likely to degrade on the target domain $\mathcal{D}_{t}$ which presents a domain shift. Among existing works, one of the key ideas is to match the \emph{input space} for both domains in the feature space so that the mapping can be invariant to the inputs. It could be achieved by adversarial training \cite{tzeng2017adversarial} or prior matching \cite{ouyang2019data}.

\begin{figure}[t]
	\begin{center}
		\includegraphics[width=0.9\textwidth,height=0.28\textwidth]{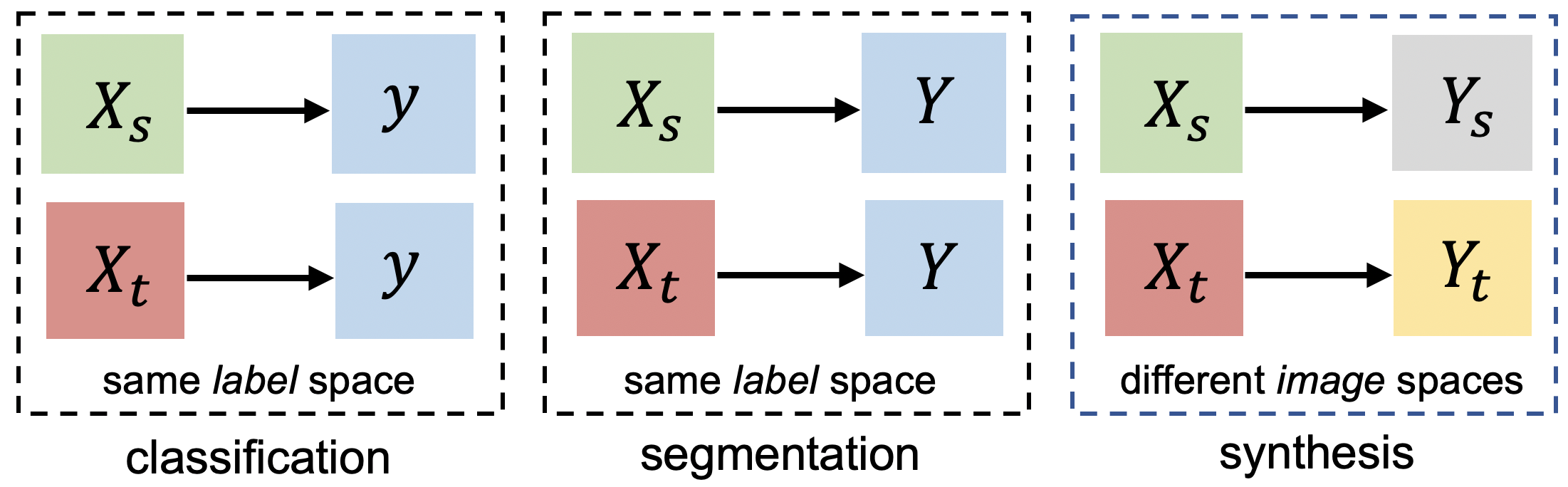}
	\end{center}
    	\caption{Summary of the main differences in domain adaptation between image classification, segmentation, and synthesis tasks. The output spaces with same colors indicate the output spaces have the same distribution. For example, a segmentation model segments the same anatomical structure in both domains.}
	\label{fig:main_idea} 
\end{figure}

As shown in Figure~\ref{fig:main_idea}, in both classification and segmentation tasks, the output label spaces in source and target domain are shared. For example, a segmentation model segments the same anatomical structure in both domains. However, in a synthesis model, the \emph{output} spaces from source domain $Y_{s}$ and target domain $Y_{t}$ are most likely different, for example, the outputs images $Y_{s}$ and $Y_{t}$ could be from different scanners. In the UDA scenario, we only have access to the input of target domain $X_{s}$, thus matching the synthetic output $\hat{Y}_{t}$ to its real distribution is challenging as there is no observations of the outputs. Importantly, aligning $X_{t}$ and $X_{s}$ does not guarantee that the output would be close to $Y_{t}$ but $Y_{s}$. Thus, most existing works in classification and segmentation could not be directly applied to synthesis model. Generally, we expect the generated output of the target domain $\hat{Y}_{t}$ to match a \emph{reasonable} distribution of the target domain. 
In this work, we present the problem setting, upper bound and propose an efficient approach to perform UDA in a 3D setting. 
\subsubsection{Why 3D-UDA is necessary and challenging?} Previous work focusing on 2D or patch-based adaptation \cite{dou2018unsupervised,unsupervised2017KK}. Although these works show promising results, they are limited to 2D or patch domains which is insufficient for many applications such as neuroimaging data which requires domain adaptation in a 3D fashion. The 3D image-to-image synthesis model dealing with full-volume imaging data is heavy-weight compared to patch-based method. However, extending existing work from 2D to 3D is non-trivial. In addition to model complexity, another challenge is that the number of 3D volumetic samples is very limited while 2D slices are more accessible.  


\subsubsection{Contributions.}
Our contribution is threefold: (1) We introduce unsupervised domain adaptation for 3D medical image synthesis and present the technical difference with existing setup in image classification and segmentation. (2) We propose an efficient 2D variational-autoencoder approach to perform UDA in a 3D manner. (3) We present empirical studies on the effects of the amount of data needed for adaptation and the effect of key hyper-parameters. Our results show that the proposed method can significantly improve the synthesis accuracy on unseen domains.


\begin{figure}
    \centering
    \includegraphics[width=0.78\textwidth]{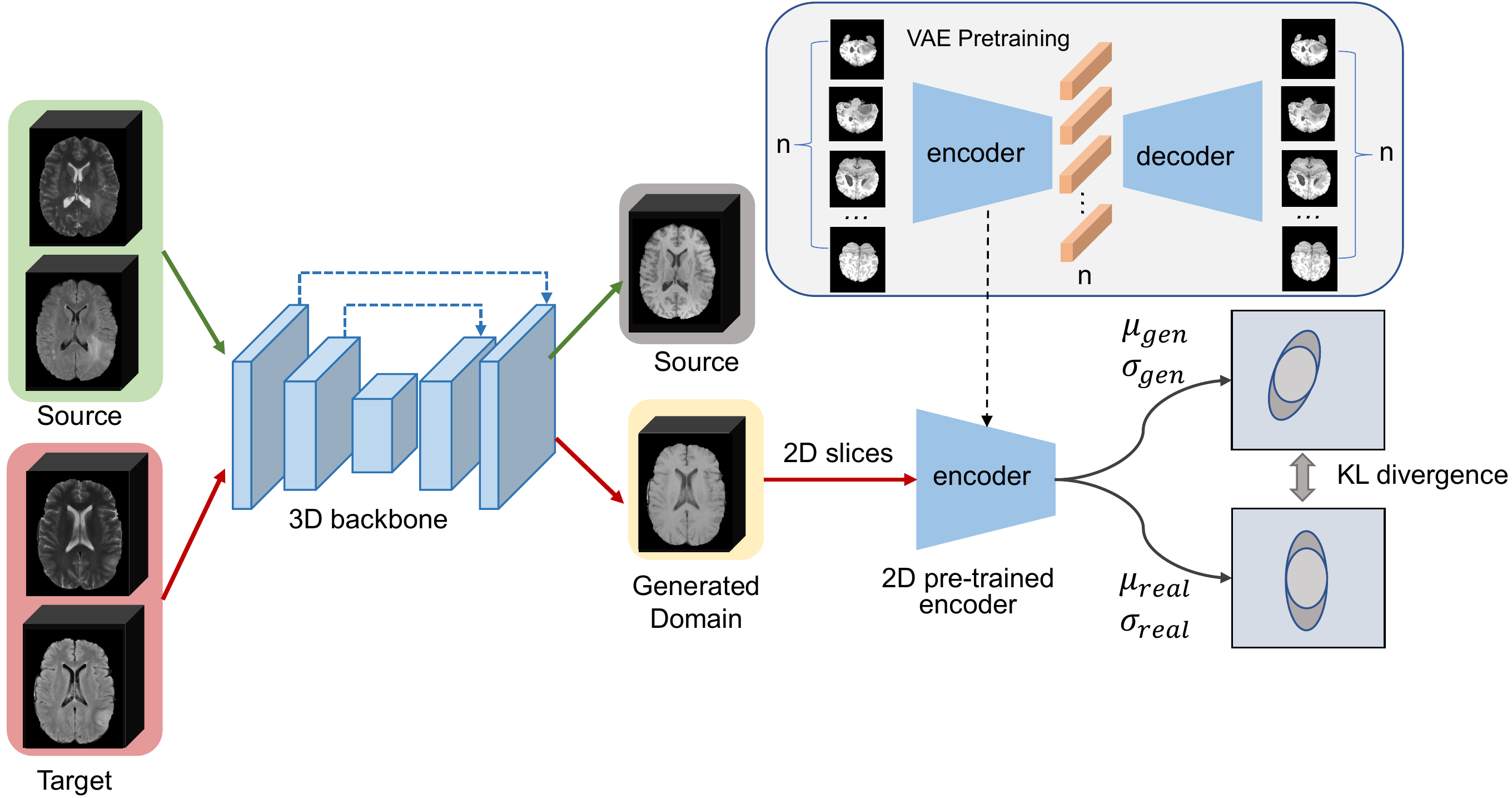}
    \caption{The main framework of our proposed 3D-UDA method for cross-modality MR image synthesis. In the source domain (green), the 3D backbone is trained supervisedly with aligned image data, translating FLAIR and T2 to T1. A spatial 2D variational autoencoder is first trained in the source domain using T1 modality. The dimension of its latent space is $n$. Then, in the adaptation stage, we compute the KL-divergence between the prior distribution of the generated volume and the target 3D distribution learned by a 2D variational autoencoder.}
    \label{fig:main_uda}
\end{figure}
\section{Methodology}
\subsubsection{Problem definition.} The objective is to adapt an volume-to-volume mapping $\Phi_{s}$: $X_s$$\rightarrow$$Y_s$ which is trained on a source domain to a target domain, so that when testing on input data $X_t$, the output could match the target distribution: Let $\mathcal{S}$ and $\mathcal{T}$ denote the source domain and the target domain, respectively.
We observe $N$ samples ${S} = \{(x_{s}^{k}, y_{s}^{k})\}_{k=1}^{N}$ from $\mathcal{S}$,
and $M$ samples ${T} = \{x_{t}^{j}\}_{j=1}^{M}$ from $\mathcal{T}$.
Notably, the samples from the target domain do not contain any output data.  
\paragraph{Supervised domain adaptation.} When there is some target data $\{X_{t}, P(X_{t})\}$ available, one could use them to fine-tune the established mapping $M$ and transfer the knowledge from source to target. When increasing the amount of data for tuning, the upper bound could be setup for \emph{unsupervised} domain adaptation in which only the input data from the target domain can be accessible. 

\paragraph{Unsupervised domain adaptation.}
In this setting, $X_t$ is available while $Y_t$ is not accessible. Since the goal of a synthesis model is to generate \emph{reasonable} output. One straightforward approach is to match the 3D prior distributions of ${\hat{Y}_t}$ and ${Y}_s$. Although ${Y}_s$ and ${Y}_t$ are expected to be different, they largely share the underlying distribution, e.g., images from different scanners may present varied contrasts but share the same space of anatomical structure. However, directly modeling 3D distribution with limited data is challenging. As an alternative, we explore to model the 3D distribution with a 2D spatial variational autoencoder (s-VAE) which is effective and computationally efficient. 

\subsubsection{2D s-VAE for modeling 3D distribution.} 
To encode 3D distribution in the 2D VAE's latent space, we proposed to train the VAE in a volumetric way, i.e., instead of training the 2D VAE with slices from different brain volumes, we take shuffled 2D slices from \emph{a whole 3D volume} as the input. Thus, the batch size corresponds to the number of slices in the volume. We nickname such a form of VAE as spatial VAE (s-VAE). Doing this ensures that the 2D s-VAE learns the correlation between 2D slices. Since each latent code comes from a specific slice of a whole 3D brain volume, $n$ latent codes with a certain sequence together can express the 3D distribution, while a standard 3D VAEs encode the distribution in their channel dimension. 2D s-VAE can reduce learnable parameters compared to 3D VAEs. 
The training of s-VAE is formulated as:
\begin{equation}
    \mathcal{L}_{VAE} = D_{KL}(N(0, I)||N(\mu_{\hat{Y}_s}, \sigma_{\hat{Y}_s})) + ||Y_s - \hat{Y}_s ||_2
\end{equation}

\subsubsection{Backbone 3D ConvNets for image synthesis.} One basic component is a 3D ConvNets backbone for image synthesis. With the $N$ paired samples from the source domain, supervised learning was conducted to establish a mapping $\Phi_{s}$ from the input image space ${X}_{s}$ to the output space ${Y}_{s}$, optimized with an L1 norm loss:
$\mathcal{L}_{syn} = \sum_{i=1}^N||x_s^{i} - x_t^{i}||_1$.

\subsubsection{3D UDA with 2D s-VAE.}
Once the 3D distribution of the output in source domain $P(Y_{s})$ is learned by 2D s-VAE, we could match it with the posterior distribution $P(\hat{Y}_{t})$ given the generated output $\hat{Y}_{t}$. 
Kullback–Leibler (KL) divergence is employed to match $P(\hat{Y}_t)$ and $P(Y_s)$, which can be formulated as 
\begin{equation}
    \mathcal{L}_{ada} = D_{KL}( P(Y_s) || P(\hat{Y}_t)) = \sum P(Y_s) log(\frac{P(Y_s)}{P(\hat{Y}_t)})
\end{equation}

However, just optimizing the KL divergence can be problematic since the mapping from input to output might suffer from catastrophic forgetting. Consequently, we perform supervised training on the source domain while adapting it to target domain with KL divergence. The whole UDA pipeline is shown in Fig.~\ref{fig:main_uda}.

\section{Experiments}
\subsubsection{Datasets and preprocessing.}
We use the multi-center \emph{BraTS} 2019 dataset \cite{menze2014multimodal} to perform cross-modality image-to-image synthesis and investigate domain adaptation. Specifically, we generate T1 images from the combination of FLAIR and T2 images.
To create source and target domains for \emph{UDA}, we split it into two subsets based on the IDs of medical centers. The first subset contains 146 paired multi-modal MR images from \emph{CBICA} while the second subset consists of 239 paired MR images from \emph{TCIA}. In the supervised image-to-image synthesis training process and domain adaptation process, input volumes loaded as gray scaled image are first cropped from (155, 240, 240) to (144, 192, 192) to save memory cost. Then three data argumentation methods are applied to the cropped volumes, including random rotation, random flip and adjusting contrast and brightness. The contrast level and brightness level are  chosen from a uniform distribution $(0.3, 1.5]$. 

In our setting, there are two domain adaptation tasks: CBICA $\rightarrow$ TCIA; TCIA $\rightarrow$ CBICA. Specifically, in the first task, the image synthesis model is trained in a supervised manner using CBICA and then the model is adapted to TCIA subset. The second task is the reverse of the adaptation direction.

In the supervised training, all data from the source domain are used; in the unsupervised domain adaptation stage, we utilize 100 input volumes (FLAIR and T2) from the target domain without any output volumes (T1). The rest of volumes from the target domain are then evaluating the performance of all methods. The details of composition of datasets are summarized in Tab.~\ref{tab:dataset_composition}.


\begin{table}[]
    \centering
    \caption{The composition of datasets for the settings of three scenarios in two domain adaptation tasks. (Adapt.=Adaptation)}
    \begin{tabular}{l | c  c c | c  c c } 
\hline
                 \multirow{2}{*}{Methods} & \multicolumn{3}{c|}{CBICA $\rightarrow$ TCIA} & \multicolumn{3}{c}{TCIA $\rightarrow$ CBICA} \\ 
                 \cline{2 - 7}
                 & \multicolumn{1}{c}{Source} & \multicolumn{2}{c|}{Target} & \multicolumn{1}{c}{Source} & \multicolumn{2}{c}{Target} \\
                 &  &~Adapt. set~ &~Test set&  & ~Adapt. set~&~Test set\\
            
\hline
Without DA & 146 & 0 & 139 & 146 & 0 & 46 \\ \hline 
 UDA &146&100 & 139 & 239 & 100 & 46\\ \hline

Supervised DA & 0 &40-100 & 139 & 0 & 40-100 & 46 \\
\hline
    \end{tabular}

    \label{tab:dataset_composition}
\end{table}

\subsubsection{Configuration of the training schedule.}
We use a 3D pix2pix model \cite{pix2pix2017} as the pipeline to perform cross-modality synthesis. The generator of the pix2pix model has a large $9 \times 9 \times 9$ receptive field and it has only four down-sampling stage. Thus, due to these two specially careful designed features, the generator performs well in synthesis task while saving plenty of memory usage. FLAIR and T2 volumes are concatenated into a two-channel input. 
For the two domains, we train the 2D s-VAE model individually using T1 volumes from each of the source domain. A single volume is re-scaled from (240, 240, 155) dimension to 256 slices with dimension (256, 256). 2D s-VAE is trained for 300 epoch and the KL divergence loss weight is set to be 0.001. For the synthesis model, we first train the model using source dataset in a supervised way. Then, in the UDA process, we train the model for five epochs. In the first iteration of the UDA process, we perform supervised training on the source domain with previous hyper-parameters; in the second iteration, we fine-tune the 3D backbone with the pre-trained 2D s-VAE model. All models are trained on RTX 3090 with Pytorch 1.10.2. Due to page limit, details of the backbone synthesis architecture, 2D VAE and 3D VAE are shown in the Appendix. 

\subsubsection{Evaluation protocol.} 
We use structural similarity index (SSIM) \cite{ssim2004wang} and peak signal-to-noise ratio (PSNR) to evaluate the image quality of the synthesized images. We further use a pre-trained nnUnet \cite{isensee2021nnu} on the \emph{BraTS} 2020 dataset \cite{menze2014multimodal} to segment the generated images from different methods and report Dice scores. All the test images are not involved in the UDA process.

\begin{table}[]
\fontsize{7.0}{7.2}\selectfont
\centering
\caption{Comparison of our methods with the lower bound, upper bound, two baselines, supervised method without domain shift and real images (only on Dice). The p-values for: a) ours vs. lower bound, and b) ours vs. 3D-VAE are all smaller than 0.0001, indicating our method outperforms the two methods significantly.} 

\begin{tabular}{l|c c c|c c c}
\hline
\multirow{2}{*}{Methods} & \multicolumn{3}{c|}{~~CBICA~~$\rightarrow$TCIA~~} & \multicolumn{3}{c}{~~TCIA~~$\rightarrow$ ~~CBICA~~} \\ 
\cline{2-7}
 &SSIM   &PSNR &Dice   &SSIM   &PSNR  &Dice  \\ \hline
 Without DA (lower-bound)~~&   0.837 ($\pm$ 0.030)   &   19.999 ($\pm$ 2.150)     & 0.793  &   0.837 ($\pm$ 0.027) & 18.976 ($\pm$ 3.685) & 0.871\\ \hline
Without DA+augmentation ~~  & 0.845 ($\pm$ 0.028) & 21.525 ($\pm$ 2.204) & 0.774  & 0.833 ($\pm$ 0.029) & 19.101 ($\pm$ 3.646) & 0.874 \\ 

3D VAE UDA  & 0.844 ($\pm$ 0.026) & 21.183 ($\pm$ 2.252) & 0.772 & 0.832 ($\pm$ 0.029) & 19.278 ($\pm$ 3.611) & 0.874 \\
2D s-VAE UDA (Ours)  &    0.853 ($\pm$ 0.024)    &    22.217 ($\pm$ 2.253)  & 0.773  &  0.846 ($\pm$ 0.024)  & 19.591 ($\pm$ 3.429)  & 0.874    \\ 
 \hline
Supervised DA ($n = 10$) &0.844 ($\pm$ 0.024) & 24.969 ($\pm$ 2.634) &  0.763 &  0.851 ($\pm$ 0.014) & 22.509 ($\pm$ 2.062) & 0.864 \\ 
Supervised DA ($n = 40$)  & 0.869 ($\pm$ 0.026) & 24.933 ($\pm$ 2.828) & 0.790 &  0.852 ($\pm$ 0.017) & 23.811 ($\pm$ 2.365) & 0.866 \\ 
Supervised DA ($n = 100$)   & 0.869 ($\pm$ 0.026) & 24.619 ($\pm$ 2.953) & 0.799 &  0.865 ($\pm$ 0.017) & 23.877 ($\pm$ 2.611)& 0.870  \\ \hline
Without Domain Shift  & 0.911 ($\pm$ 0.0263) & 25.519 ($\pm$ 3.630) & 0.820  & 0.896 ($\pm$ 0.020) & 24.656 ($\pm$ 2.907) & 0.867 \\ \hline 
Real Images & - & - & 0.904 & - & - & 0.936 \\ \hline 
\end{tabular}
\label{tab:baseline_results}
\end{table}

\section{Results}
\subsection{Comparison of methods}
We first present the lower bound of the synthesis model without DA on the target domain. Then we present our proposed 2D s-VAE method. We also present the upper bound of a supervised DA.  Quantitative results are summarized in Tab.~\ref{tab:baseline_results}. Moreover, we study the impact of the amount of volumes used in the UDA process and the impact of different batch sizes in 2D VAE reconstruction results, showed in Fig.~\ref{fig:vae_samples_results}. 
\subsubsection{Lower-bound and upper-bound.} The lower-bound of the UDA is pre-training on source domain and directly testing on the target domain. Notably basic data argumentation like random rotation rotation, random flipping are used to prevent models from over-fitting. As we can observe from row 1 of Tab.~\ref{tab:baseline_results}. It achieves the worst performance of all the methods.
Given available paired data from the target domain, one could tune the pre-trained model (same as transfer learning) to adapt the model to the target domain. One could observe from the last three rows in  Tab.~\ref{tab:baseline_results}.  

\subsubsection{Heuristic data argumentation.} Heuristic data argumentation could potentially improve the generalizability \cite{zhang2020generalizing}. We perform contrast and brightness-related data augmentation considering that one of the most important domain shift is the image contrast between different scanners. We observe slight improvement by comparing row 1 and row 2 in Tab.~\ref{tab:baseline_results}. 

\subsubsection{3D VAE \emph{vs.} 2D s-VAE.} As another baseline to be compared with our proposed efficient 2D s-VAE approach, we trained 3D VAEs using the volumetic data from the source domains. 
One could observe that the 3D VAE performs comparably with the heuristic data argumentation approach. This is partly because there are limited data to train the 3D VAE for learning a proper underlying distribution. 

Our proposed 2D s-VAE method outperforms both data augmentation and the 3D VAE method on both SSIM and PSNR in two tasks. 3D VAE encoder is more computationally expensive, since the encoder of 3D VAE has 5.17M learnable parameters while 2D s-VAE only has 1.73M ones. Although there is still a visible performance gap between all the UDA methods and the upper bound, our 2D s-VAE method provides an effective and efficient solution when the output modality from the target domain is not accessible. 



\begin{figure}
    \centering
    \includegraphics[width=0.95\textwidth]{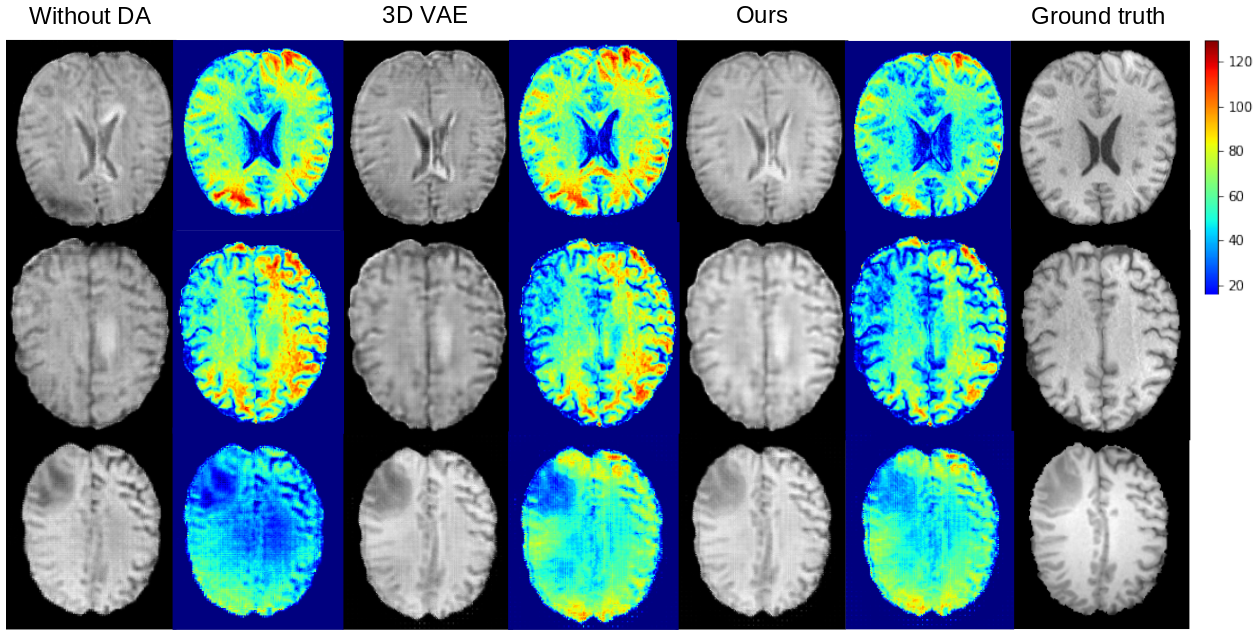}
    \caption{Results on example slices achieved by different methods: (a) Without domain adaptation (DA), (b) 3D VAE, and (c) our 2D s-VAE. The difference map (in heatmap) is computed by subtracting the ground truth slice and the synthesized slice. We observe that our approach achieves best perceptual quality.}
    \label{fig:results_comparison}
\end{figure}

\subsection{Analysis of parameters.}

\begin{figure}
    \centering
    \includegraphics[width = 0.95\textwidth]{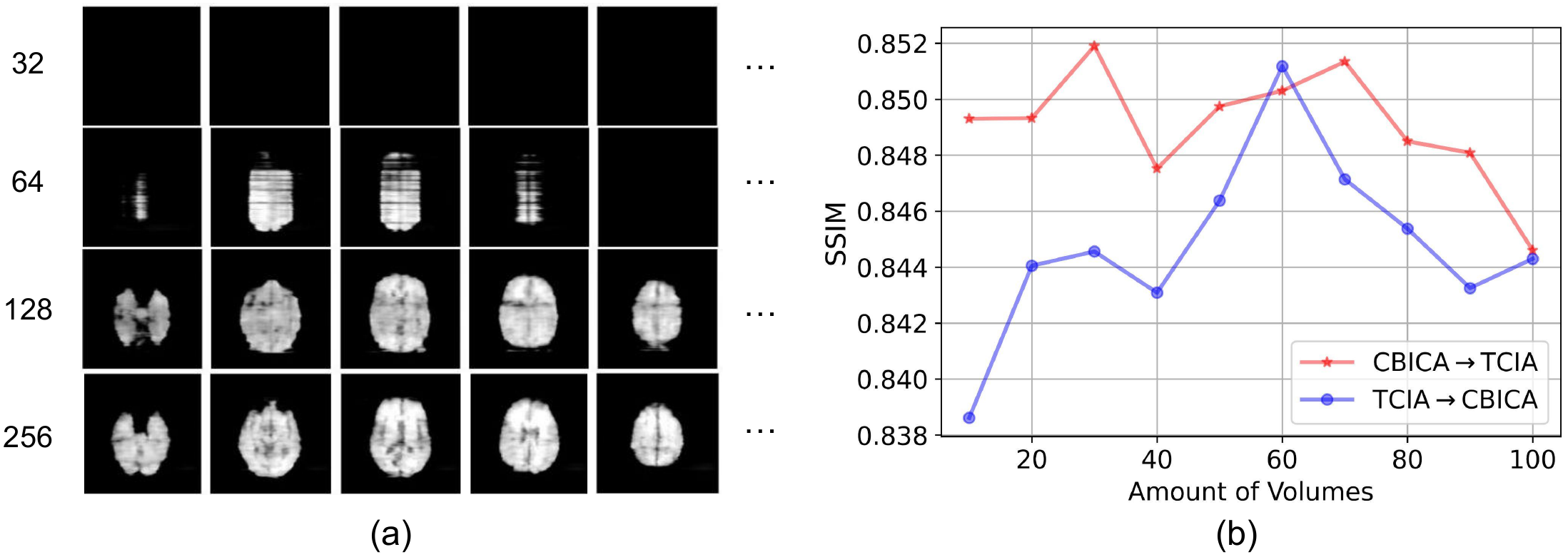}
    \caption{(a): the reconstruction results influenced by different batch sizes. Larger batch size could better capture the 3D distribution. (b): the effect of the amount of volumes used for UDA from the target domain.}
    \label{fig:vae_samples_results}
\end{figure}

\paragraph{Impact of batch size on 2D s-VAE}:
In 2D s-VAE training process and the UDA process, slices of a whole brain MRI sequence serve as the input. To show the impact of batch size on 2D s-VAE, we explore the batch size of values, 32, 64, 128 and 256. To understand how much the VAE models the 3D distribution and visualize the results from the learned prior, we follow \cite{volokitin2020modelling} to build a Guassian model of 2D slices, which models the correlation of 2D slices in the latent space for sampling infinite number of 2D slices. As we show in Fig.~\ref{fig:vae_samples_results} (b), when batch size is 32, almost nothing can be sampled from the learned distribution; when batch size is 64, only noise can be sampled; when batch size is 128 (at this point half of the brain slices are used in a batch), brain shape can be visualized in the samples; when batch size is 256, brain structures are clear in the samples.
\paragraph{Impact of the amount of volumes}:
As we show in Fig.~\ref{fig:vae_samples_results}(b), we study the impact of amount of volumes used for the UDA process. We observe that for both $\text{CBICA} \rightarrow \text{TCIA}$ and $\text{TCIA} \rightarrow \text{CBICA}$ tasks, when the number of volumes is less than 70, the performance increases. However, when the number exceeds 70, the performance starts to decrease. That is because the first continuing training batch in the UDA process contributes more to the results rather than the second batch. Although the first batch regulates the whole UDA process, it might hurt the performance in some degree.


\section{Conclusion and discussion}
In this work, for the first time, we explore domain adaptation for medical image-to-image synthesis models. We first explain the difference between the domain adaptation of synthesis, classification and segmentation models. Then we introduce our efficient unsupervised domain adaptation method using 2D s-VAE when the target domain is not accessible. Finally, we show the effectiveness of our 2D s-VAE method and study some factors that influence the performance of our framework. In our approach, we translate the whole volume from one domain to another instead of using a patch-based method. Although whole-volume approaches are able to capture the full spatial information, it suffers from limited training data issues. As we have shown in Fig.~\ref{fig:results_comparison}, even after domain adaptation, we observed that the domain gap is challenging to overcome. Recent disentangled learning that could separate domain-specific and shared features effectively might improve the current results. Contrastive learning could be explored to better capture the representation of the source or target domains more effectively. Given the above limitations, we still wish our approach provides a new perspective for robust medical image synthesis for future research. 



\bibliographystyle{splncs04}
\bibliography{egbib}

\appendix
\section{Visualization of network backbones}


\begin{figure}
    \centering
    \includegraphics[width=0.95\textwidth]{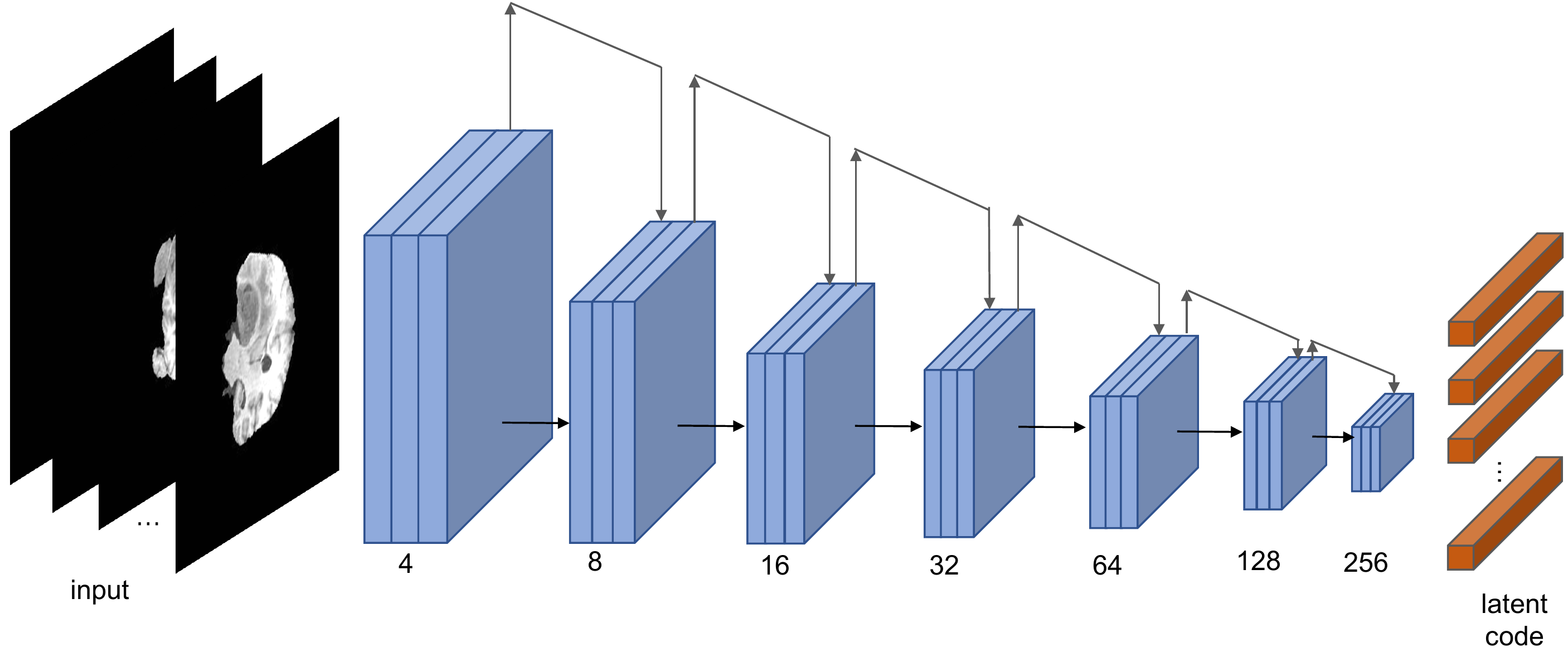}
         \caption{The encoder structure of the 2D s-VAE. The building blocks of the encoder is  $3\times 3$ ResNet convolutional blocks. Each input channel dimension is shown in the figure. 2d s-VAE encodes the 3D spatial information in its batch size dimension. }
    \label{fig:2d_encoder}
\end{figure}

\begin{figure}
    \centering
    \includegraphics[width=0.95\textwidth]{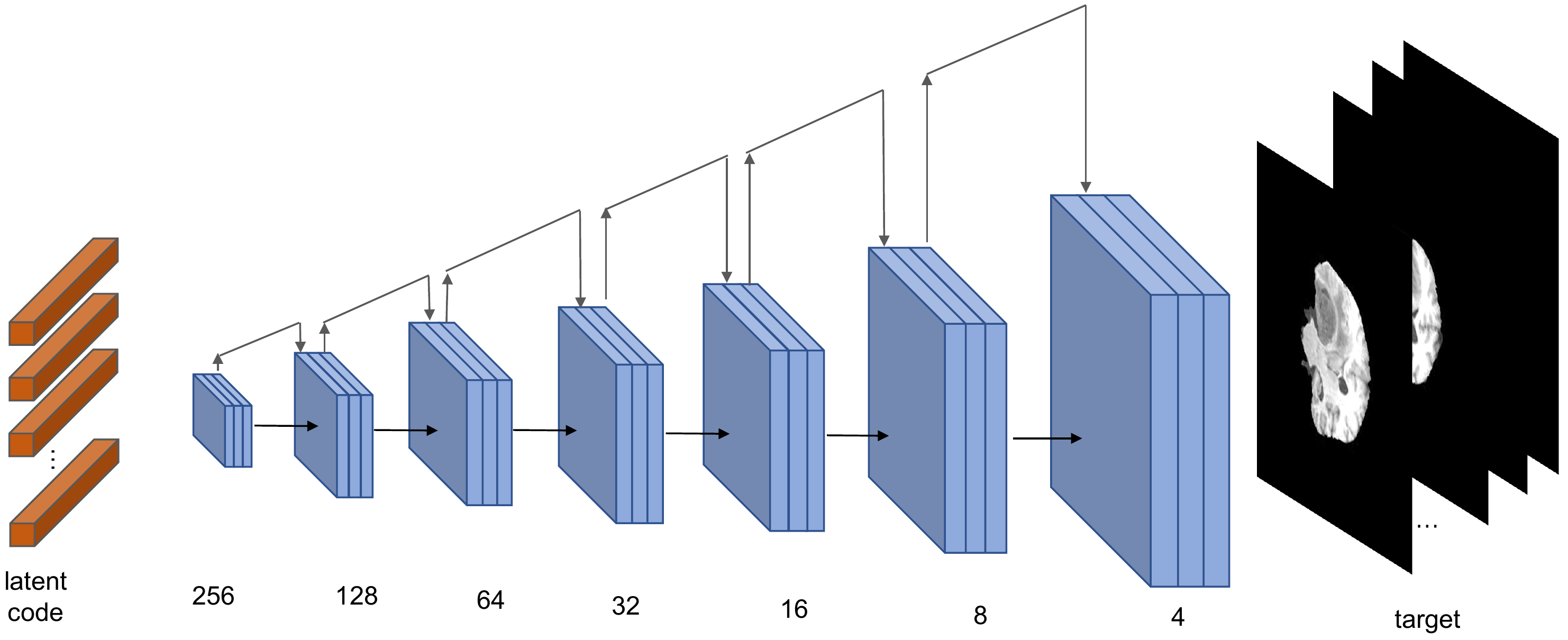}
         \caption{The decoder structure of the 2D s-VAE. The building blocks of the decoder are also $3 \times 3$ ResNet blocks with upsampling operation.}
         \label{fig:2d_decoder}
\end{figure}
\begin{figure}
    \centering
    \includegraphics[width=0.95\textwidth]{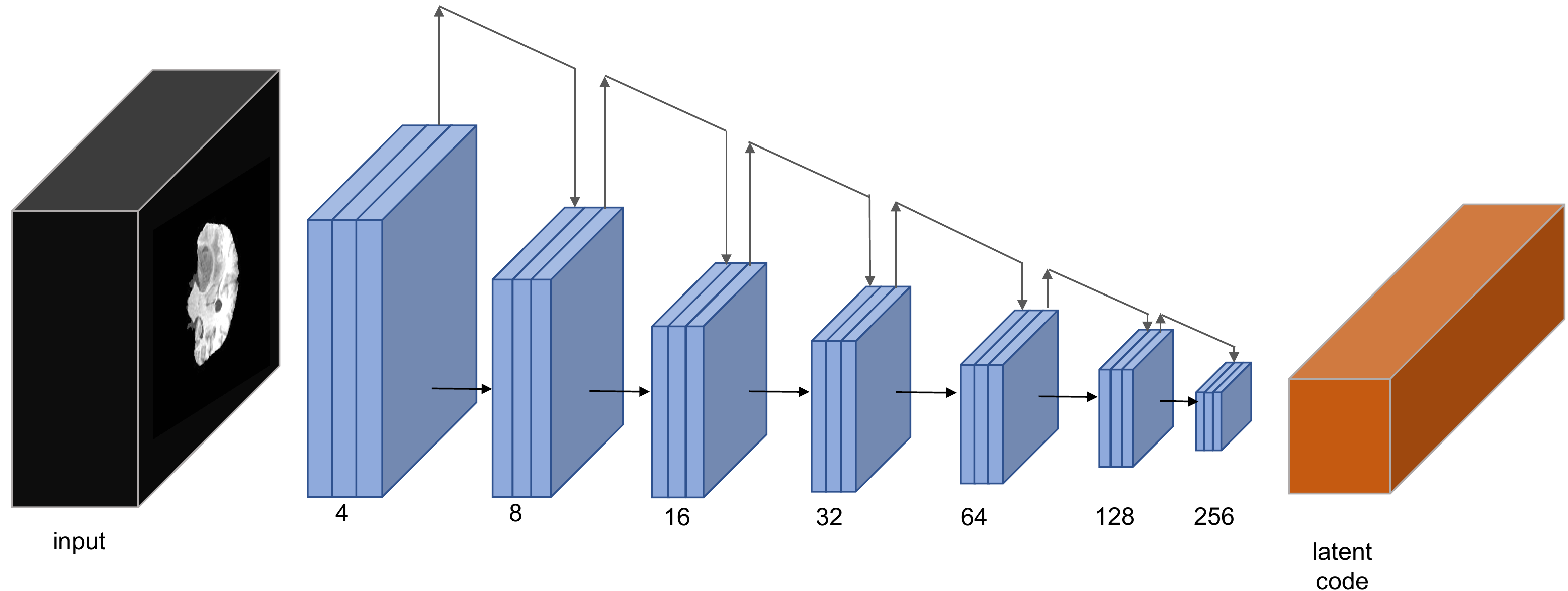}
         \caption{The encoder structure of the 3D VAE. The structure is similar to the 2D s-VAE, but 3D VAE encodes the 3D spatial information in its channel dimension.}
    \label{fig:3d_encoder}
\end{figure}

\begin{figure}
    \centering
    \includegraphics[width=0.95\textwidth]{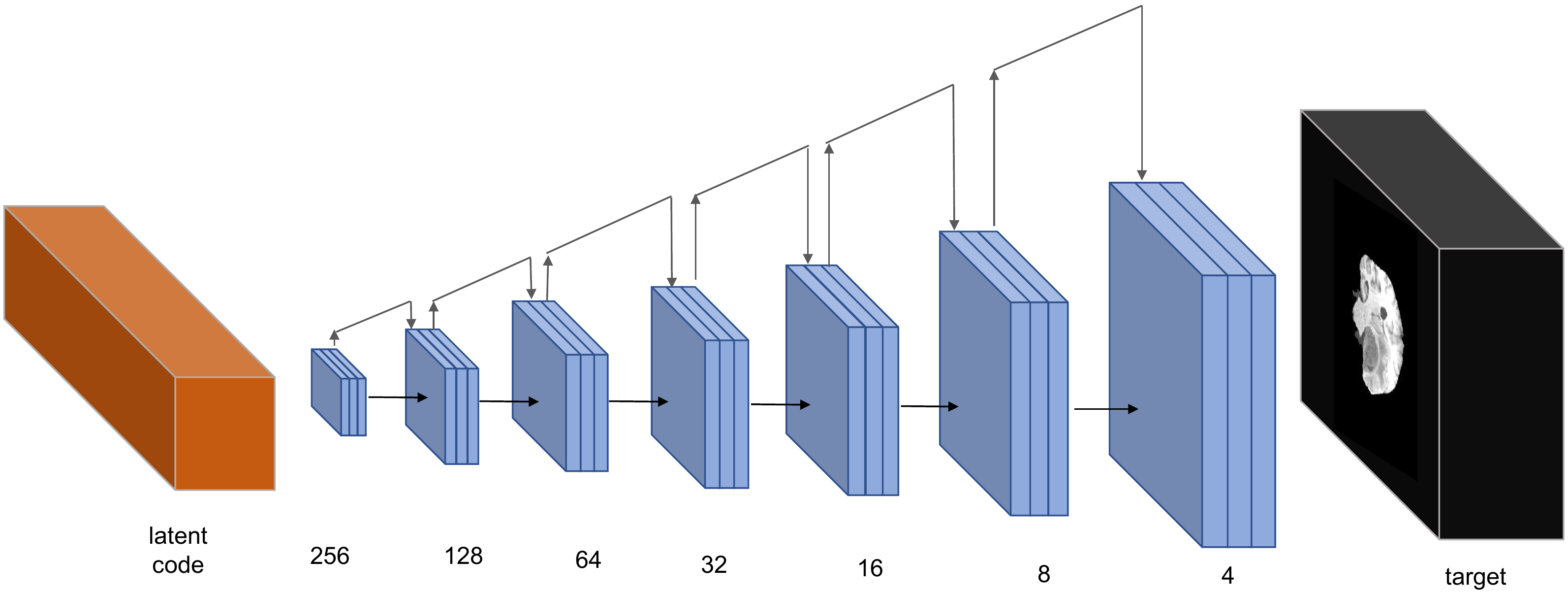}
         \caption{The decoder structure of the 3D VAE.}
         \label{fig:3d_decoder}
\end{figure}

\end{document}